\documentclass[12pt]{article} \textheight 23cm
\usepackage{graphicx}
\usepackage{amsmath}
\usepackage{color}
\begin{document} \begin{center}
{\large \bf Oscillatory Spreading and Surface Instability of a Non-Newtonian Fluid under Compression }\\ \vskip 0.5cm
Moutushi Dutta Choudhury$^1$, Subrata Chandra$^1$, Soma Nag$^1$, Shantanu Das$^2$
and Sujata Tarafdar$^1$\\
\vskip 0.5cm

$^1$ Condensed Matter Physics Research Centre, Physics Department, Jadavpur University, Kolkata 700032, India\\
$^2$ Reactor Control Division, Bhabha Atomic Research Center, Trombay, Mumbai 400085, India\\

\end{center}
\vskip 1cm
\noindent {\bf Abstract}\\ Starch solutions, which are strongly non-Newtonian, show a surface instability, when subjected to a load. A droplet of the fluid is sandwiched between two glass plates and a weight varying from 1 to 5 kgs. is placed on the top plate. The area of contact between the fluid and plate increases in an oscillatory manner, unlike Newtonian fluids in a similar situation. The periphery moreover, develops a viscous fingering like instability, which is not expected under compression. We attempt to model the non-Newtonian nature of the fluid through  a visco-elastic model incorporating generalized calculus. This is shown to exhibit a qualitatively similar oscillatory variation in the surface strain.
\vskip 0.5cm
\noindent

Forced spreading of a fluid under an impressed force is an interesting topic \cite{Engmann,bonn,nag}. The physics involved is challenging and the problem has important applications in technology as well. In real life many fluids are non-Newtonian, adding further complexity to the problem.

In the present work we report studies on spreading of a starch solution between two glass plates, when the upper plate is loaded by a weight. We observe an interesting oscillation in the area of contact between the fluid and glass plate as a function of time. Earlier study of Newtonian fluids \cite{nag} did not show such behavior. We try to explain this phenomenon using  fractional calculus, which is known to be an appropriate technique to study non-Newtonian, visco-elastic materials \cite{visco,das1,das2}. Another remarkable observation is the appearance of a surface instability similar to viscous fingering. Viscous fingering under the condition of lifting, i.e. separating the plates is a well studied phenomenon \cite{epje,martine,gay}. In this case the pressure is lower within the fluid, compared to the air pressure outside, satisfying the Saffman-Taylor condition for instability \cite{saffman}. However, in the present case the fingering develops during {\it compression}.

\begin{figure}[ht]
\begin{center}
\includegraphics[width=14.0cm, angle=270]{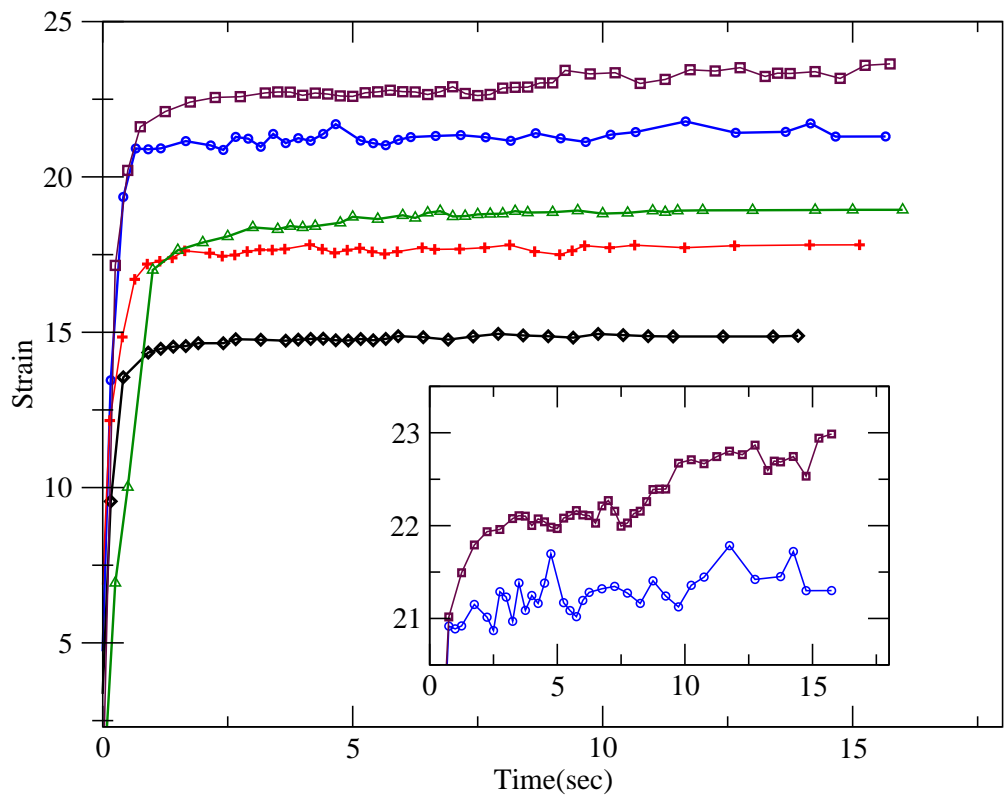}
\end{center}
\caption{Experimental variation in strain with time for different values of load for 2.5 concentration arrowroot solution on glass. The inset shows the oscillations at higher magnification. Symbols  - diamond (black), plus (red), triangle (green), circle (blue) and square (violet) represent respectively the effects of loads 1,2,3,4 and 5 kgs. in addition to the glass plate which weighs around 0.6 kgs.} \label{expt}
\end{figure}

\begin{figure}[ht]
\begin{center}
\includegraphics[width=14.0cm, angle=270]{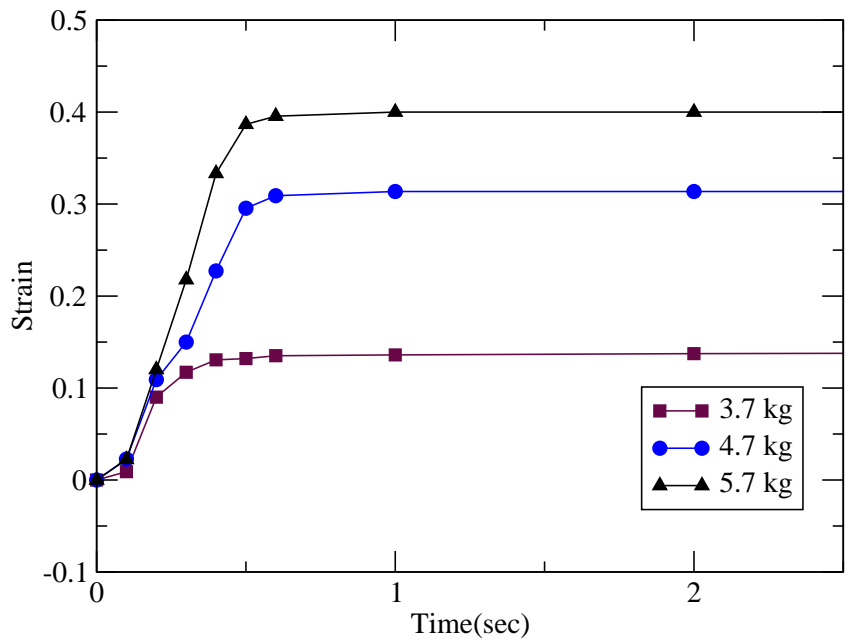}
\end{center}
\caption{Variation in strain with time for different values of load for ethylene glycol on glass. } \label{newtn}
\end{figure}

\begin{figure}[ht]
\begin{center}
\includegraphics[width=14.0cm, angle=0]{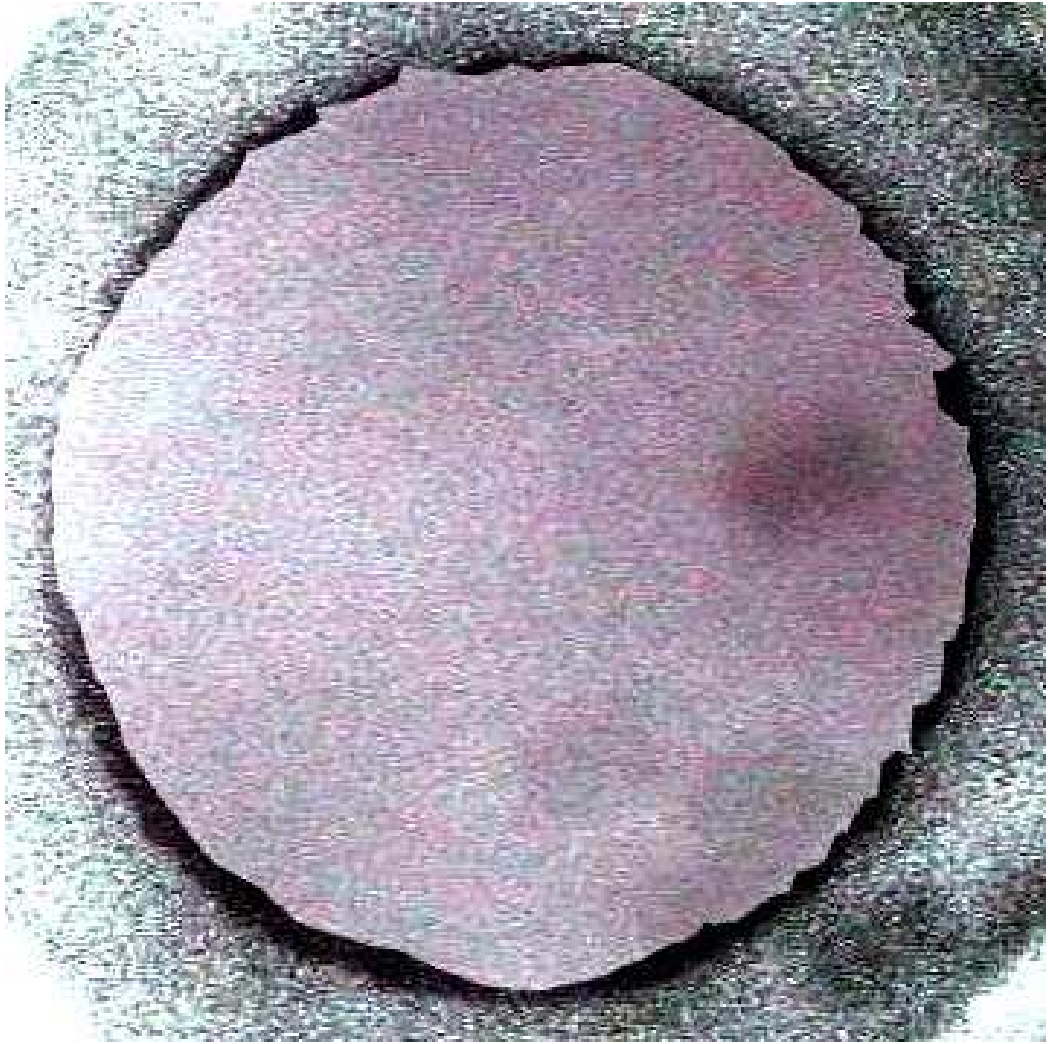}
\end{center}
\caption{A snapshot of the film (inner light colored blob) superposed on the photograph of the film photographed 1 second earlier (darker outline visible along the periphery) shows the shrinking of the film.} \label{outline}
\end{figure}

The fluid under study is prepared by adding $m$ gms of arrowroot to 100 ml distilled water and boiling for 1 minute making the solution thicken. Here $m$ = 2.5 and 3.5 gms. A pinch of dye is added to enhance contrast. A drop of solution is placed on a glass plate and another glass plate is placed on top. The upper plate is loaded by a weight $W$, which varies from 1 to 5 kgs. The spreading drop is photographed from below using a video camera. The fractional change in the area of contact of fluid and glass is measured using Image-Pro Plus software and plotted as a function of time in Figure \ref{expt}. A similar figure for ethylene glycol, which is Newtonian is shown for comparison in Figure \ref{newtn}. An oscillatory behavior is clearly visible for the starch solution. For the load of 1kg. the variation is nearly smooth, but oscillations are quite pronounced for 4 and 5 kgs. as shown more clearly in the inset. It is to be noted that this is not simply stick-slip behavior, where the strain would always increase, but in jumps. In this case, the strain actually decreases, before increasing again. To establish that the film {\it does} shrink and expand periodically, we show a superposition of two snapshots in figure(\ref{outline}) taken at an interval of 1 sec. Here the larger outer boundary corresponds to an instant of time earlier than the inner. So the ups and downs in figure(\ref{expt}) are genuine and not due to measurement error. Further, the corrugated appearance of the boundary demonstrates the instability developed.

\begin{figure}[ht]
\begin{center}
\includegraphics[width=14.0cm, angle=270]{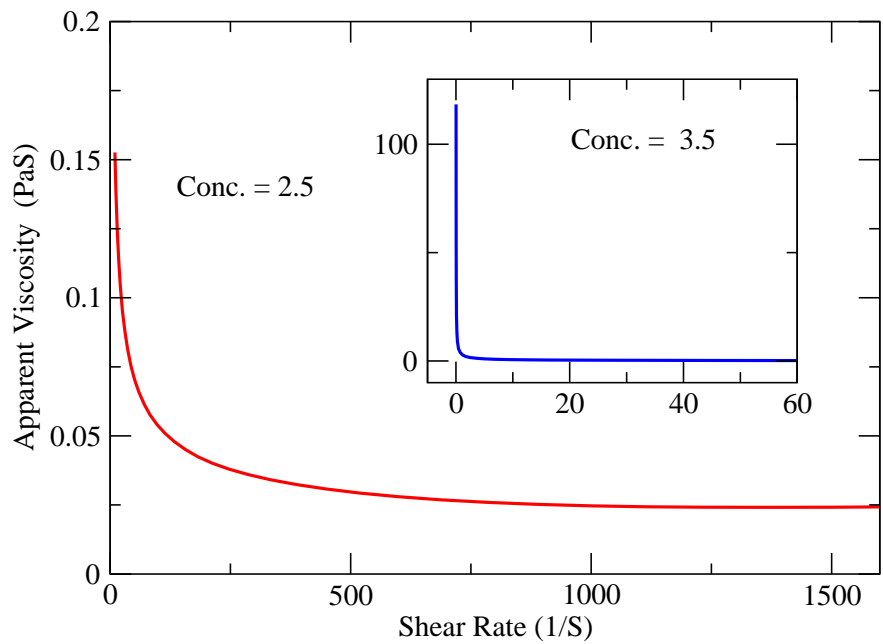}
\end{center}
\caption{Variation of apparent viscosity with strain rate for the arrowroot solutions shows the non-Newtonian behavior, which is stronger for higher concentration (inset). } \label{rheo}
\end{figure}

Rheological study of the two starch solutions was done at the Central Glass and Ceramic Research Institute, Kolkata by a Bohlin rheometer. Viscosity plotted against the rate of strain in Figure (\ref{rheo}) reveals the non-Newtonian nature of the fluids, which is more prominent for the higher concentration, as shown in the inset. This is in agreement with earlier results \cite{moorthy}.

 To analyze the stress-strain behavior here, we use the standard Maxwell model \cite{max} of a spring and dashpot in series. The viscous term represented by the dashpot is further generalized by taking the $q$th derivative of the strain. Here $q$ may be a fraction, for case  $q$ = 1 we have a Newtonian viscosity. The elastic term is assumed to be Hookean at present.
The stress is then given by 

\begin{equation}
 \sigma = \beta \frac{d^q\epsilon}{dt^q} + E \epsilon
 \label{basic}
\end{equation}
Here $\epsilon$ is the strain, $E$ the elastic modulus and $\beta$ a parameter characterizing the effective viscosity of the non-Newtonian fluid.

We assume a step function to represent the loading $$ \sigma(t) = \sigma \text{ for } t\ge 0  \text{, and  }  \sigma(t) = 0 \text{ for }  t<0 $$.
The initial condition for strain is $\epsilon(t) = 0$ for $t<0 $.

The Laplace transform of equation(\ref{basic}) gives

\begin{equation}
 \epsilon(s) = \frac{\sigma}{E} \left[\frac{1}{s} - \frac{s^{q-1}}{s^q + E/\beta}\right]
 \label{lap}
\end{equation}
The inverse Laplace transform of equation(\ref{lap}) gives
\begin{equation}
 \epsilon(t) = \frac{\sigma}{E}\left[1 - ML_q\bigg(-\frac{E}{\beta} {t^q}\bigg)\right]
\label{strain}\end{equation}

where, ML(-kt) is the one parameter Mittag-Leffler function defined by
$$ ML_1(-kt) = e^{-kt}$$ $$ ML_q(z) = \sum_{k=0}^\infty \frac{z^k}{\Gamma(qk+1)} $$

\begin{figure}[ht]
\begin{center}
\includegraphics[width=14.0cm, angle=270]{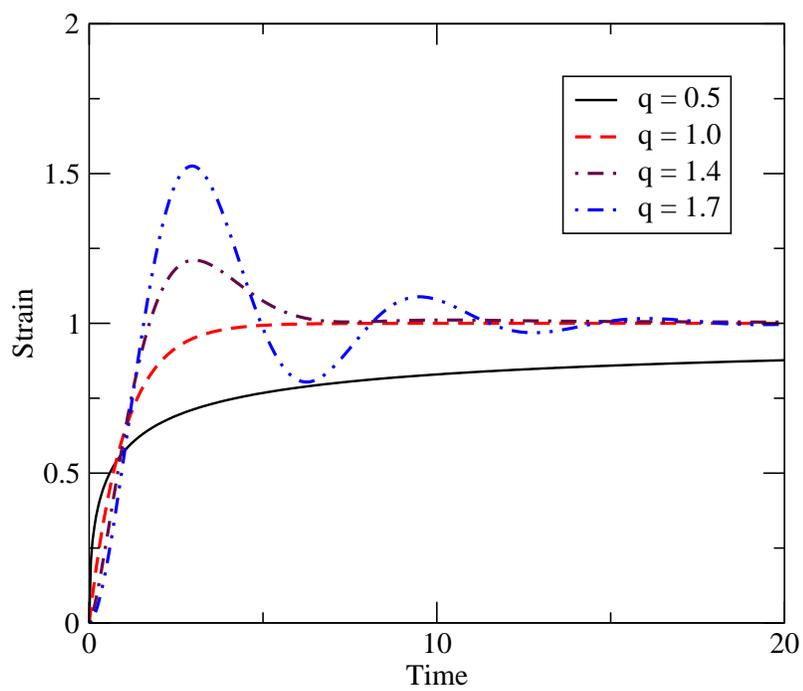}
\end{center}
\caption{Variation in strain with time for different values of $q$. } \label{qvar}
\end{figure}

We now plot the strain as a function of time using equation(\ref{strain}). For $q$ = 1, the strain increases smoothly and saturates to the value $\epsilon_\infty = \sigma/E$. For $q < 1$, a similar behavior is observed, but the saturation value is less. For $q > 1$ however, we see an initial increase in strain overshooting $\epsilon_\infty$ followed by oscillations before saturating to $\epsilon_\infty$. The oscillations are more pronounced as $q$ increases. These results are shown in figure(\ref{qvar}). 

The equation \ref{basic} is the generalized representation of the stress-strain relation. Let us rewrite the same as:

\begin{equation}
 \frac{d^q\epsilon(t)}{dt^q} + B \epsilon(t) = \frac{1}{\beta} \sigma(t)
\end{equation}

Where $B = E/\beta$. Here $\beta$  is the generalized viscosity coefficient with units corresponding to the 
non-integer order. When the order $q = 1$, then normal coefficient of viscosity is recovered. The unit of B for order $q = 1$
is per seconds i.e. $s^{-1}$ , but for any other order $q \ne 1$; the unit modifies to $s^{-q}$.

Mathematically one has to see the Green's function for general relaxation in equation\ref{basic}, so we write the homogeneous
equation with RHS equal to zero. To that, we give delta function stress excitation. The strain built up for any relaxation
process may be treated as convolution integral of a strain variable with integral kernel  ${K_q}(t)$, as \cite{das1,das2}.

\begin{equation}
 \frac{d}{dt}\epsilon(t) = -\int_0^t K_q{(t-\tau)}\epsilon(t) dt
\end{equation}

Well if the memory kernel is $K(t) = B_0 \delta(t)$, we have the above system \ref{basic} without memory \cite{das1,das2} and the Green's function will be

$$ \epsilon(t) = \epsilon_0 e^{-B_0t} $$
that is the impulse response quickly decays to zero. Here $\epsilon_0$ is initial strain of the system at $t = 0$.
 
 This can be derived as follows:
 
 \begin{equation}
 K(t) = B_0 \delta(t)
\end{equation}

\begin{equation}
\frac{d}{dt}\epsilon(t) = -\int_0^t \delta(t-\tau)\epsilon(t) dt = -B_0\epsilon(t)
\end{equation}

\begin{equation}
\epsilon(t) = \epsilon_0 e^{-B_0t} 
\end{equation}

The homogeneous strain relaxation equation for no-memory case is a first order Ordinary Differential Equation i.e.

\begin{equation}
 \frac{d}{dt}\epsilon(t)+B_0\epsilon(t) = 0
\end{equation}

If the memory kernel is a constant say $K_2(t) = B_2$, then we will have oscillatory Green's function, which never decays to zero \cite{das1,das2}.

\begin{equation}
 K_2(t) = B_2
\end{equation}

\begin{equation}
 \frac{d^2}{dt^2}\epsilon(t) = -B_2\epsilon(t)
\end{equation}

\begin{equation}
 \epsilon(t)=\epsilon_0 cos(\sqrt{B_2}t)
\end{equation}

The generalized memory integral is \cite{das1,das2}.

\begin{equation}
 K(t)=B_qt^{q-2} \quad ; \quad 0<q\le 2
\end{equation}

\begin{equation}
 \frac{d}{dt}\epsilon(t)=-\frac{1}{\tau^q}\bigg[\frac{d^{(1-q)}}{dt^{(1-q)}}\epsilon(t)\bigg]
\end{equation}

\begin{equation}
 \tau^q = \big[B_q \Gamma(q-1)\big]^{-1}
\end{equation}

Its corresponding generalized differential equation, obtained from above derivation, is the system with memory with the
 memory index coming as fractional order of the Fractional Differential Equation with, $0<q\le 2$.
 
 \begin{equation}
 \frac{d^q\epsilon(t)}{dt^q}-\epsilon_0\frac{t^{-q}}{\Gamma(1-q)} = -\tau ^{-q} \epsilon(t) \label{fr} 
\end{equation}

In our experiment, to describe the oscillatory response to a step input we say that the order is between  $1<q<2$ and thus the system has a long lingering memory.
In equation(\ref{fr}) above, if the  initial stress be  $\epsilon_0$ , assuming Heaviside's step function as the stress input, it modifies to
equation(\ref{basic}), for $\tau^{-q} \rightarrow B$.

The order of the equation(\ref{basic}) corresponds to a system with memory. The non-Newtonian fluids without oscillatory behavior will have $0<q<1$, which is
fractional order, and the step-response will have monotonically increasing strain response, given by one argument
Mittag-Leffler function. Its impulse response will be having long tailed decay. That is the response will have long-range
temporal correlation. The Newtonian fluid will have integer order in equation(\ref{basic}) with $q=1$, representing the system without memory, and the
step-response will have monotonically increasing strain as
 $\epsilon(t) \sim 1-exp(B_0t)$;
 where its impulse response will decay quickly as $\epsilon(t) \sim exp(B_0t)$.
This visco-elastic system with Newtonian viscous bahavior can be modeled with a discrete ideal spring and a ideal dashpot. Whereas the more complicated case with non-Newtonian viscosity
requires a different representation like a fractal chain of the ideal spring and ideal dashpot combination. We have observed oscillatory
strain and thus infer the fractional order $q$ of our system to lie between 1 and 2.

\begin{figure}[ht]
\begin{center}
\includegraphics[width=14.0cm, angle=270]{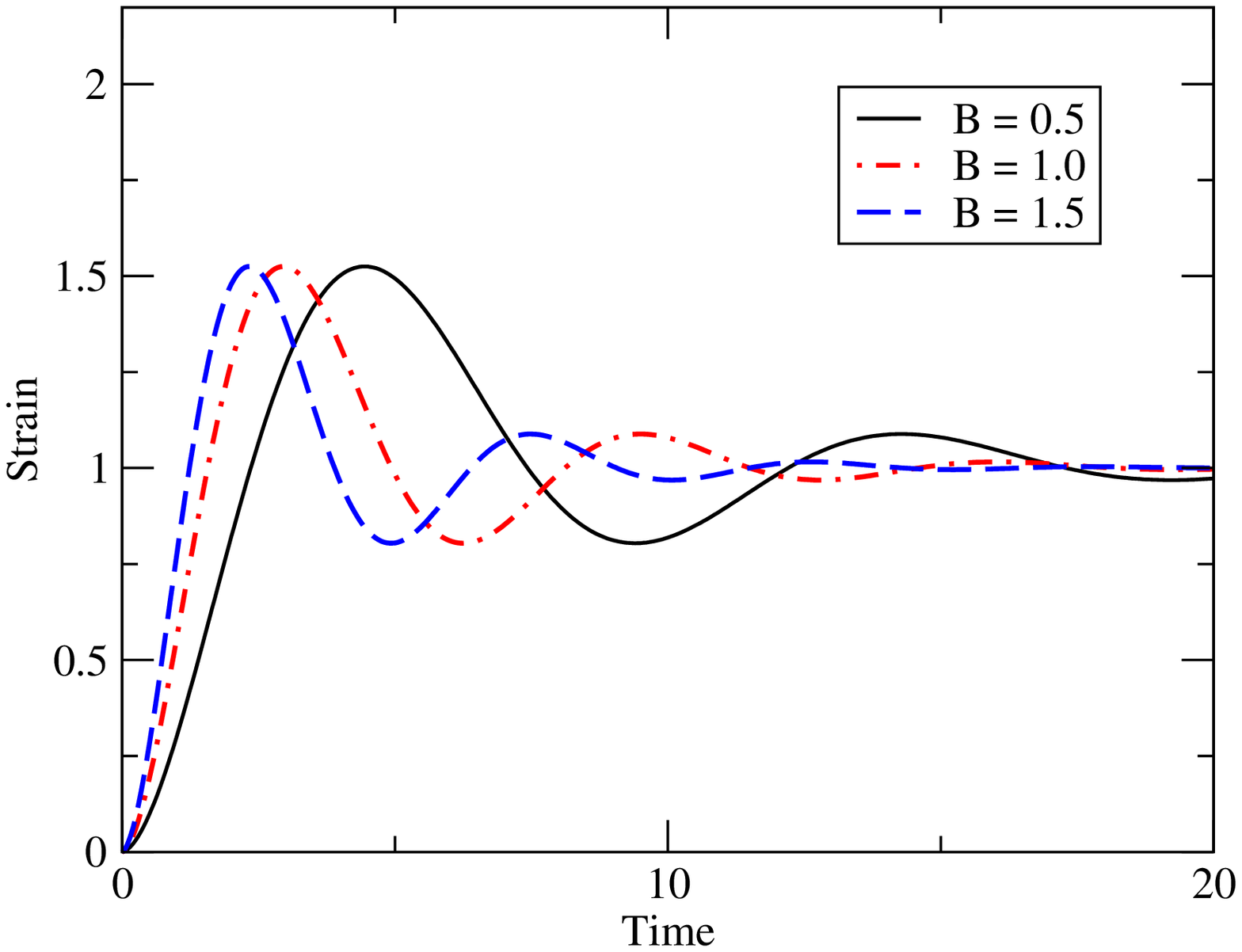}
\end{center}
\caption{Variation in strain with time for different values the parameter $B$ for $q$ = 1.7. } \label{oscB}
\end{figure}

Considering that  $B = \frac{E}{\beta}$ is a parameter representing the relative strengths of the elastic and viscous terms in equation(\ref{basic}), we may see how the system responds to changes in $B$. We find that variation of $B$ changes the time period of the oscillations in strain, without affecting the amplitude as shown in figure(\ref{oscB}). The time period is smaller when elasticity dominates. The amplitude depends on $q$ and of course the magnitude of the strain changes proportionately to the load $\sigma$, also affecting the amplitude of oscillation. 

If we assume that the fractional change in area is equivalent to the strain in the system, the results for the starch solutions are reproduced qualitatively by the visco-elastic model with non-Newtonian rheology. Here the qth order derivative takes care of the non-linearity in the complex fluid. In the earlier paper \cite{nag} the Newtonian fluids were assumed incompressible and the change in film thickness was calculated from the constant volume of the fluid. Here the fluid may have a finite compressibility, so we refer only to the area which is measured directly. 

To conclude, this work demonstrates the interesting phenomenon of oscillatory  spreading of starch solutions on glass and illustrates further how the approach of generalized calculus may be used to analyze it. The qualitative agreement of the variation in strain with time from the experiments and the theory is encouraging and further exploration along these lines promises to be rewarding.

The authors thank UGC, Govt. of India for supporting this work and for providing a research grant to MDC. Tapati Dutta and Prof. S.P. Moulik are gratefully acknowledged for helpful discussion.

\begin{thebibliography}{99}
\bibitem{Engmann}J. Engmann, C. Servais, A.S. Burbridge, Squeeze flow theory and applications to rheometry: A review, J. Non-Newt. Fluid Mech. 132(2005)1-27
\bibitem{bonn}  D. Bonn, J. Eggers, J. Meunier, E. Rolley, Wetting and spreading, Rev. Mod. Phys. {\bf 81} (2009) 739
\bibitem{nag} S Nag, S Dutta, S Tarafdar, Applied Surface Science 256 (2009) 353-355
\bibitem{visco} N. Heymans, J.C. Bauwens, Rheologica Acta, {\bf 33} (1994) 210
\bibitem{das1} S. Das, {\it Mathematico-Physics of Generalized Calculus}, (Limited edition, available at Jadavpur University and University of Calcutta, 2010)
\bibitem{das2}S. Das, {\it Functional Fractional Calculus for System Identification and Control} (Springer-Verlag, Berlin, Heidelberg, 2007)
\bibitem{epje} S. Sinha, T. Dutta, S. Tarafdar, Adhesion and fingering in the lifting Hele-Shaw cell, Eur. Phys. J. E 25(2008) 267-275
\bibitem{martine} M.Ben Amar and D.Bonn, Physica D, {\bf 209}, 1 (2005)
\bibitem{gay} C.Gay and L.Leibler, Physics Today, (Nov.1999).
\bibitem{saffman}  T.Vicsek, {\it Fractal Growth Processes} (World Scientific,Singapore,1989).
\bibitem{moorthy} S.N. Moorthy, Starch/Starke, {\bf 54} (2002) 559
\bibitem{max} R.M. Christensen, {\it Theory of Viscoelasticity} (2nd ed, Dover, New York, 2003)


\end{thebibliography}
\end{document}